\documentclass[10pt]{iopart} 

\usepackage{iopams}  

\usepackage{graphicx}
\usepackage{upgreek}
\usepackage{xcolor}

\begin{document}

\title[Laser-lithographically written micron-wide SNSPD]{Laser-lithographically written micron-wide superconducting nanowire single-photon detectors}

\author{Maximilian Protte$^{1,*}$, Varun B. Verma$^2$, Jan Philipp H{\"o}pker$^{1}$, Richard P. Mirin$^2$, Sae Woo Nam$^2$, Tim J. Bartley$^1$}

\address{$^1$ Mesoscopic Quantum Optics, Department of Physics, Paderborn University, Warburger Str. 100, 33098 Paderborn, Germany \\
$^2$ National Institute of Standards and Technology, 325 Broadway, Boulder, Colorado 80305, USA}
\ead{$^*$maximilian.protte@upb.de}
\vspace{10pt}
\begin{indented}
\item[]December 2021
\end{indented}

\begin{abstract}
We demonstrate the fabrication of micron-wide tungsten silicide superconducting nanowire single-photon detectors on a silicon substrate using laser lithography. We show saturated internal detection efficiencies with wire widths from 0.59\,$\upmu$m to 1.43\,$\upmu$m under illumination at 1550\,nm. We demonstrate both straight wires, as well as meandered structures. Single-photon sensitivity is shown in devices up to 4\,mm in length. Laser-lithographically written devices allow for fast and easy structuring of large areas while maintaining a saturated internal efficiency for wire width around 1\,$\upmu$m. 
\end{abstract}

%
%
%
%
\ioptwocol

\section{Introduction} \label{introduction}
Superconducting nanowire single-photon detectors (SNSPDs) have had a great impact on modern applied optics experiments, if not revolutionized the field of quantum optics~\cite{You2020}. SNSPDs are used in quantum communication~\cite{Yin2017}, quantum computation~\cite{Zhong2020}, as well as light detection and ranging (LIDAR)~\cite{Taylor2019} and deep-space communication~\cite{Grein2015}. These applications have generally high requirements for the detectors, and detector properties have been highly optimized. Remarkable results including detection efficiencies close to unity~\cite{Reddy2020}, extremely low dark counts~\cite{Hochberg2019}, and timing jitter in the single-picosecond range~\cite{Korzh2020} demonstrate how SNSPDs have outperformed other detector types. Nevertheless, all of these come with the drawback of high requirements for cryogenic temperatures and fabrication tolerances when structuring the superconducting thin films. The latter is especially crucial when investigating complex detector geometries~\cite{Hu2020}, large-scale detectors~\cite{Steinhauer2021}, and multi-element devices~\cite{Allman2015,Wollman2019}.

Recent demonstrations of micron-wide superconducting nanowire single-photon detectors with different superconducting materials such as NbN~\cite{Korneeva2018,Vodolazov2020,Xu2021}, MoSi~\cite{Charaev2020,Korneeva2020,Lita2021}, and WSi~\cite{Chiles2020,Wollman2021} have improved the insight into the fundamental working principle of SNSPDs. 
In contrast to conventional SNSPD structures, the dimensions of which necessitate e-beam lithography, Chiles et al.~\cite{Chiles2020} and Lita et al.~\cite{Lita2021} were able to use optical mask lithography to structure detectors with saturated internal efficiency, and Wollman et al.~\cite{Wollman2021} to demonstrate devices with a large active area.

In this paper, we report on our investigations of WSi micron-wide superconducting nanowire single-photon detectors with saturated internal efficiency, structured only with optical laser-lithography for different wire widths. This structuring process enables fast and easy prototyping of arbitrary detector geometries, "all-in-one-step" structuring of detectors, additional electronics, and contact pads, and fabrication of large-scale devices with reduced stitching of write-fields. In Section~\ref{Fabrication}, we present the fabrication process of the WSi-detectors from 400\,nm to 1.4\,$\upmu$m and its fabrication limits. Section~\ref{Results} contains measurement results of the saturated detection efficiency at 775\,nm wavelength as well as 1550\,nm wavelength under flood illumination, as well as investigations of dark counts and timing jitter, before concluding in Section~\ref{Conclusion}.

\section{Fabrication} \label{Fabrication}
\begin{figure}[htbp]
  \centering
  \includegraphics[width=0.49\textwidth]{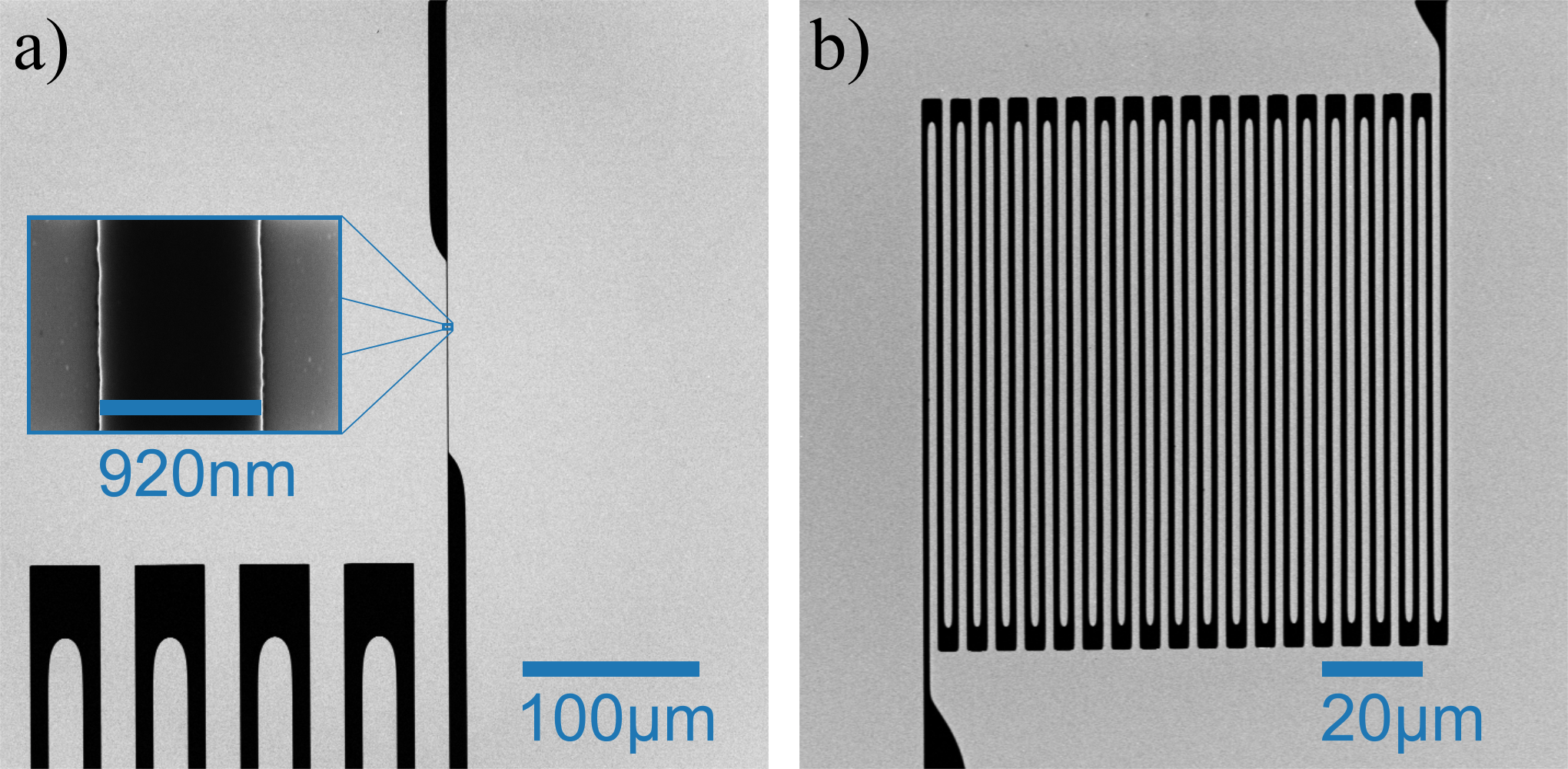} 
\caption{SEM image of a) a 920\,nm wide detector wire with serial inductor and b) a meander design.}
\label{Fig1}
\end{figure}

The fabrication process starts with magnetron sputtering of the a 2.6\,nm thick tungsten silicide film on a silicon substrate as described in Reference~\cite{Chiles2020}. The laser lithography tool for structuring the film uses a wavelength of 375\,nm, has a maximum write field of 30\,$\upmu$m x 200\,mm without stitching and a write speed of 2\,mm$^{2}$/s. After structuring and subsequent development of the photo resist, the resulting structures are dry etched with C$_{4}$F$_{8}$/SF$_{6}$. With this process, we realized SNSPDs in different geometries: straight lines from 100\,$\upmu$m length to 6\,mm length with varying width between 400\,nm and 1.4\,$\upmu$m as well as meandered shapes with a wire width of 1\,$\upmu$m and an unfolded length of 4\,mm. For detector geometries of less than 4000~squares, an additional 10\,$\upmu$m wide 5000~square inductor as well as contact pads for all tested geometries were designed using the PHIDL GDS-layout code~\cite{McCaughan2021} and written in the same writing step as the SNSPDs. Figure~\ref{Fig1} shows scanning electron microscope images (SEM) of two of the investigated detector designs.

\begin{figure}[htbp]
  \centering
  \includegraphics[width=0.45\textwidth]{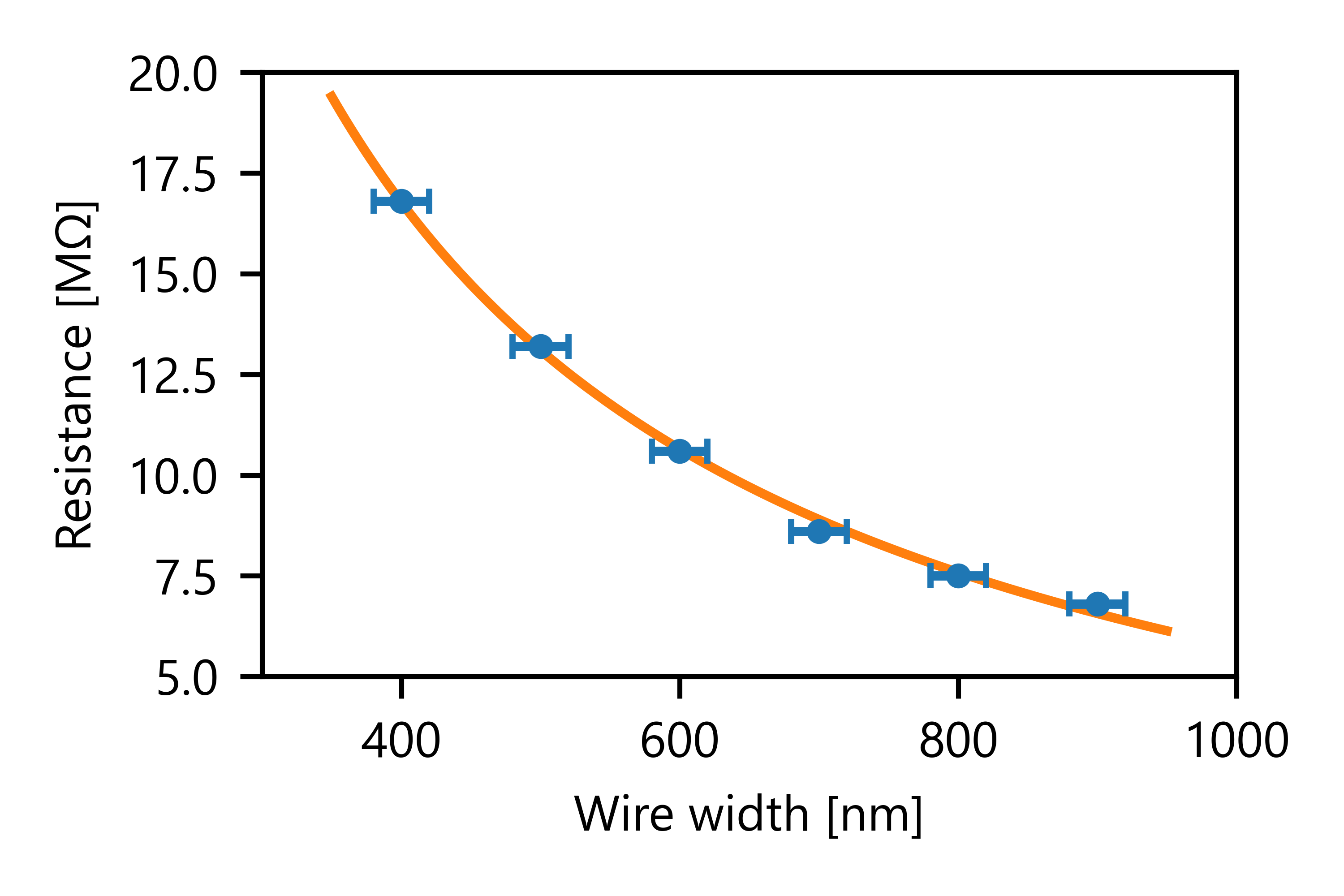} 
\caption{Room-temperature resistance of 6\,mm long detector wires and contact pads for different wire widths between 400\,nm and 900\,nm. The inversely proportional relation is illustrated by the fit.}
\label{Fig2}
\end{figure}

To investigate artefacts in the writing process, the room temperature resistance of 6\,mm long stripes with contact pads made out of the same tungsten silicide layer is measured for different wire widths, from 400\,nm to 900\,nm, as shown in Figure~\ref{Fig2}. While electron-microscope images show a slight increase in sidewall roughness for wire widths at the writing-resolution limit of 400\,nm, the measurement data shows the inversely proportional relation of wire width and room-temperature resistance.

\section{Results} \label{Results}
\begin{figure}[htbp]
  \centering
  \includegraphics[width=0.45\textwidth]{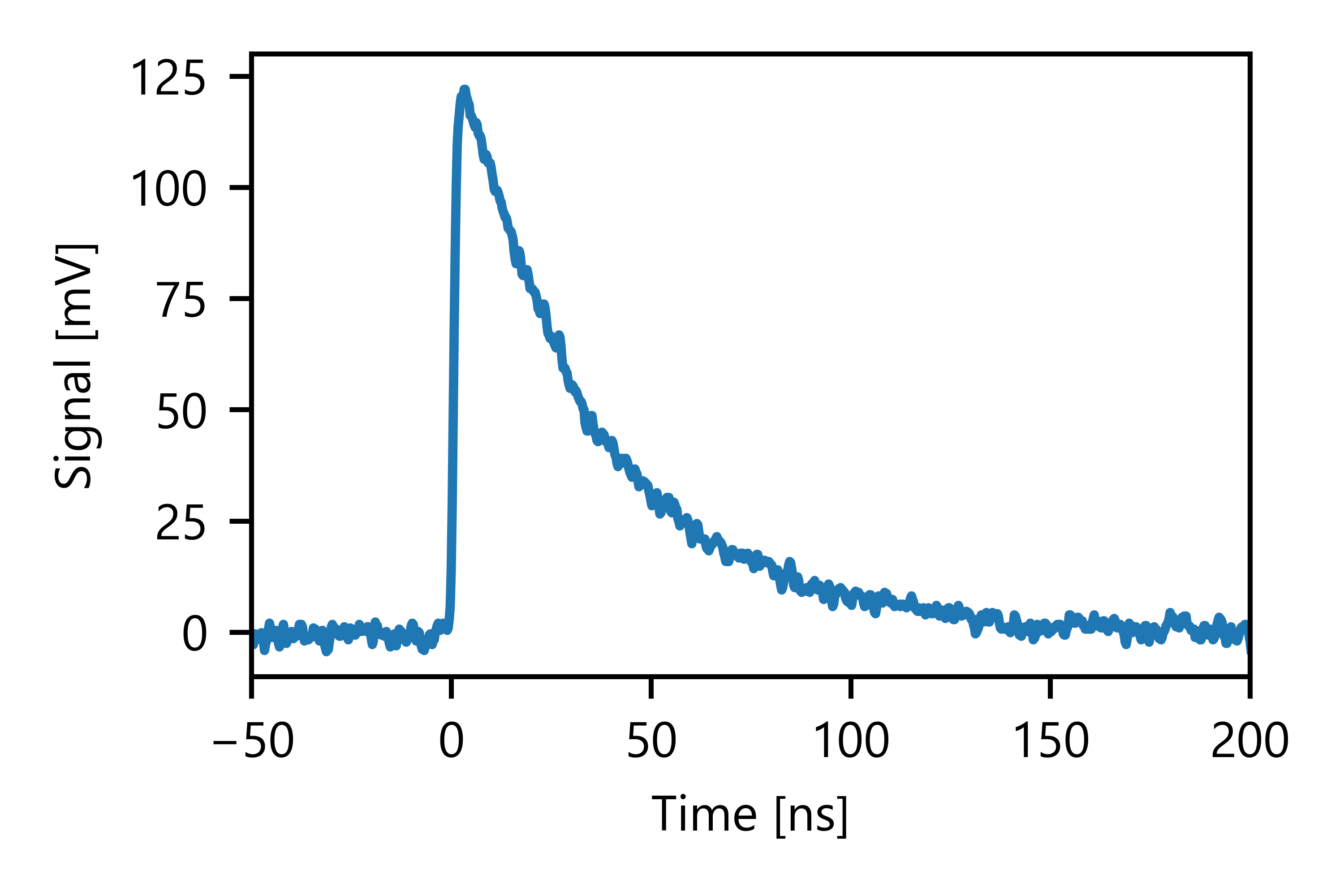} 
\caption{Electrical detector response to impinging photons of a 600\,nm wide and 200\,$\upmu$m long wire with serial inductor at 22\,$\upmu$A bias current after room-temperature amplification.}
\label{Fig3}
\end{figure}

For cryogenic investigation of the different detector geometries, several devices were wire-bonded and cooled down in a closed-cycle sorption cryostat with a base temperature of 0.8\,K. An SMF-28-Fiber at a distance of 2\,mm from the fiber end to the sample surface is used for targeted flood illumination, using an attenuated pulsed lasers of 1550\,nm wavelength and 500kHz repetition rate. Using this setup, the electrical responses to photon absorption of the different detectors are measured. All detectors with widths between 400\,nm and 1.4\,$\upmu$m and lengths between 200\,$\upmu$m and 6\,mm responded to single photons. The detector response of a 920\,nm wide and 200\,$\upmu$m long wire with a serial inductor at 22\,$\upmu$A bias current is shown in Figure~\ref{Fig3}.

\begin{figure}[htbp]
  \centering
  \includegraphics[width=0.45\textwidth]{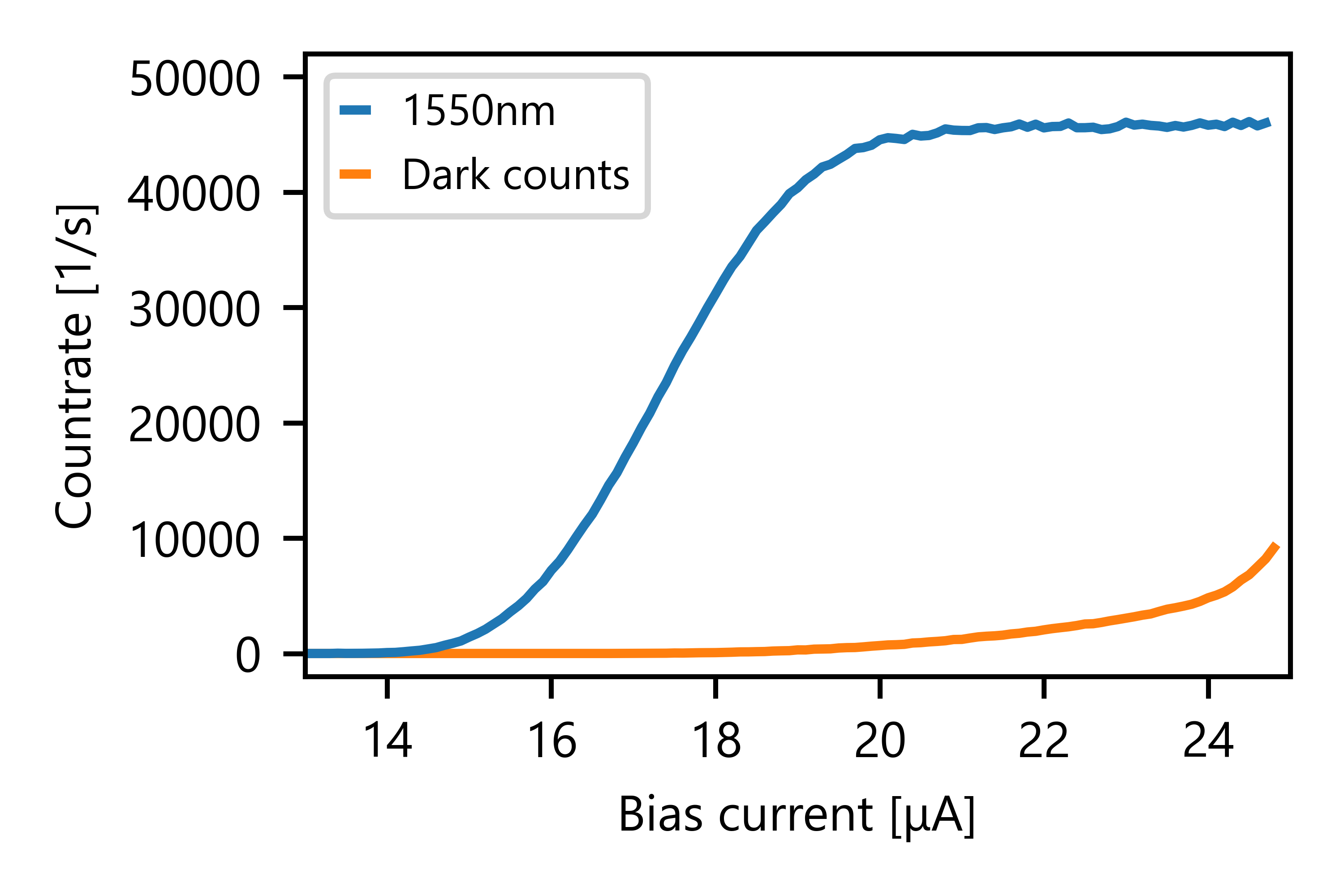} 
\caption{Bias-dependent countrate of detection events per second for the 920\,nm wide and 200\,$\upmu$m long wire, measured in coincidence with the arrival time of the laser pulse at 1550\,nm wavelength as well as detected dark counts.}
\label{Fig4}
\end{figure}

A decisive characteristic of SNSPDs is a saturated internal efficiency. This depends on the applied bias current, as only a sufficiently high bias current together with an absorbed photon of certain energy can cause a measurable detection event. For the 920\,nm wide and 200\,$\upmu$m long wire with serial inductor, the saturation of the detection efficiency is shown in Figure~\ref{Fig4}. The plateau indicates sufficient fabrication tolerances using laser lithography. As the detector responses were measured in coincidence with the pulsed laser, no additional dark-count subtraction was executed, as the number of dark counts was negligible in a total of 1\,ms active measurement time. Therefore, the shown dark-count data was taken independently.

\begin{figure}[htbp]
  \centering
  \includegraphics[width=0.45\textwidth]{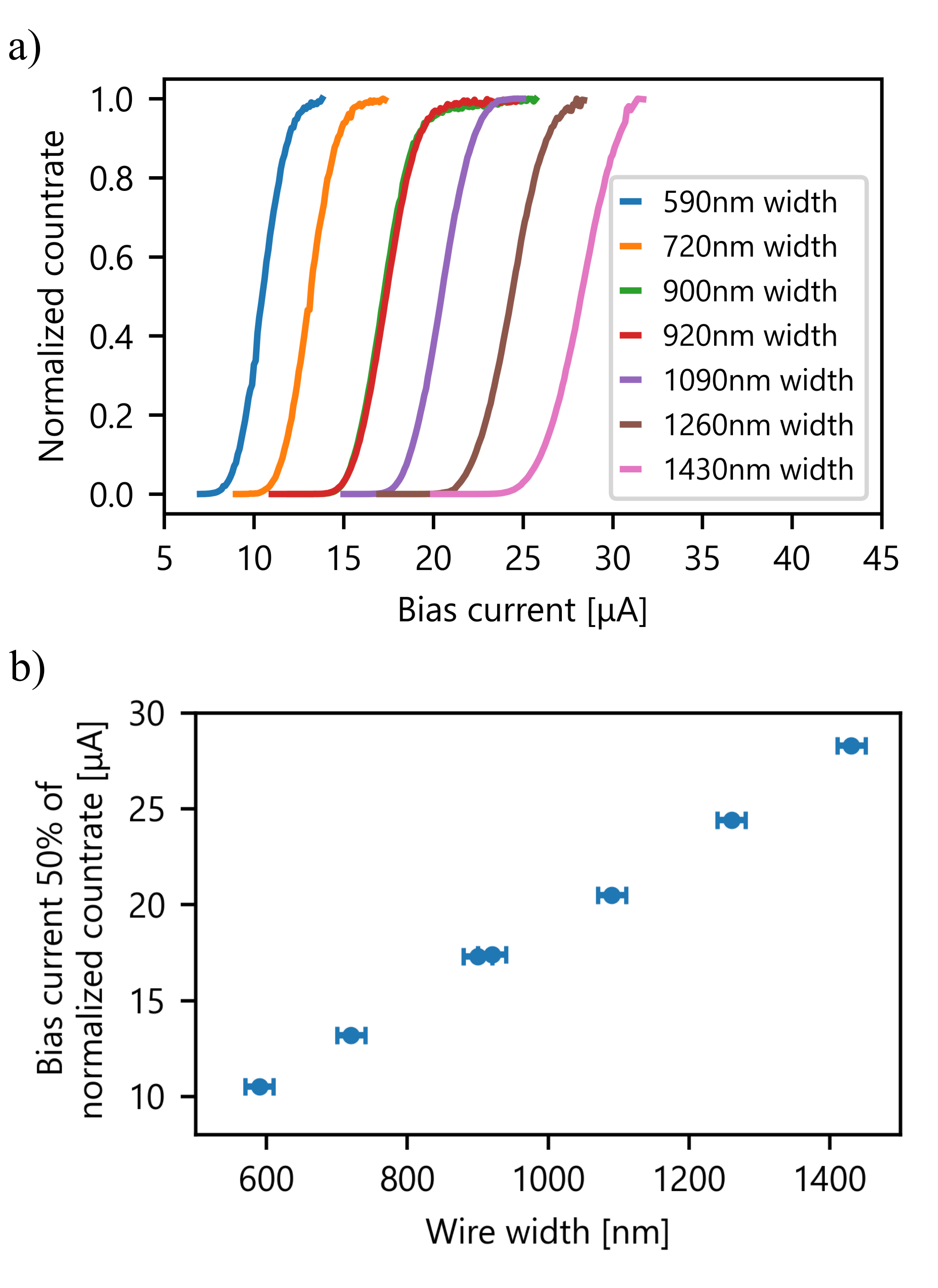} 
\caption{a) Bias-dependent countrate of detection events per second at 1550\,nm wavelength for different wire widths, measured in coincidence with the arrival time of the laser pulse and normalized to their maximum countrate before latching. The red graph for a 920\,nm wide device corresponds to the measurement data in Figure~\ref{Fig4}. b) Bias current values at 50\% of the maximum countrate based on the data from Figure~\ref{Fig5}~a) for different detector-wire widths. The wire width and displayed error bars are based on SEM width measurements.}
\label{Fig5}
\end{figure}

Furthermore, the results for similar devices with wire widths from 590\,nm to 1430\,nm are presented in Figure~\ref{Fig5}. The absolute countrates of each detector varies due to different collection efficiencies caused by the different active areas as well as the relative fiber position. 

\begin{figure}[htbp]
  \centering
  \includegraphics[width=0.45\textwidth]{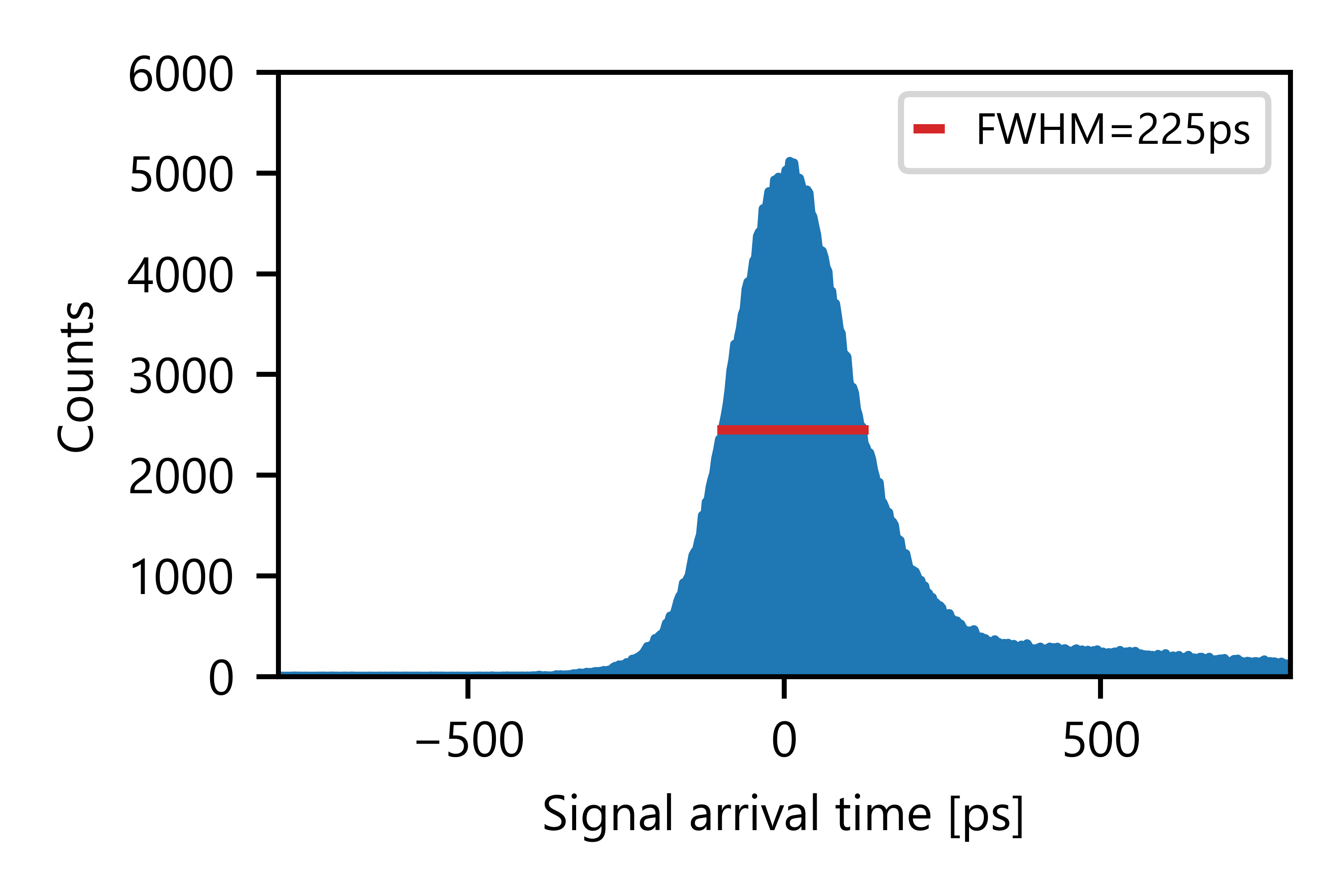} 
\caption{Jitter measurement of the 920\,nm wide and 200\,$\upmu$m long tungsten silicide wire with serial inductor for 1550\,nm photons, showing a full width at half maximum of 225\,ps.}
\label{Fig6}
\end{figure}

In addition, we measured a timing jitter of 225\,ps for the 920\,nm wide detector, as shown in Figure~\ref{Fig6}, using the pulsed laser at 1550\,nm wavelength with a pulse width of 9\,ps. Jitter contributions from 5\,mm long wire bonds, the laser synchronisation, or time tagger were not deconvoluted.

\begin{figure}[htbp]
  \centering
  \includegraphics[width=0.45\textwidth]{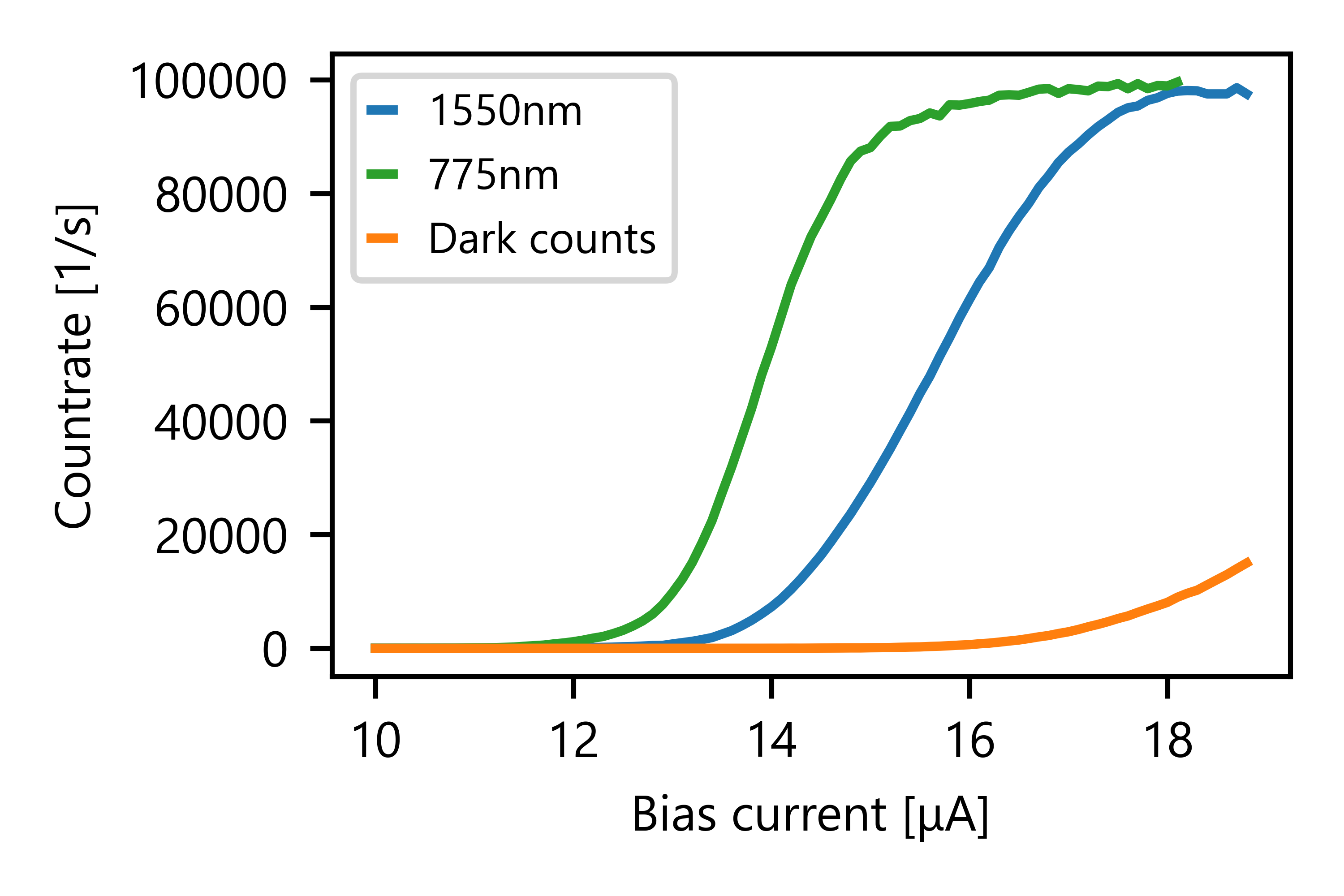} 
\caption{Bias-dependent countrate of detection events per second for the 1\,nm wide and 4\,mm long meander using the pulsed lasers at 1550\,nm and 775\,nm wavelength, measured in coincidence with the arrival time of the laser pulse as well as detected dark counts.}
\label{Fig7}
\end{figure}

As imperfections in the fabrication steps are particularly critical for large-area devices, we also investigated the bias-dependence of the detection efficiency for a 1\,$\upmu$m wide and 4\,mm long detector folded into a meander shape, as depicted in Figure~\ref{Fig1} b) for 1550\,nm photons as well as 775\,nm photons. This devices also shows a plateau region for both wavelengths, as shown in Figure~\ref{Fig7}. Deviations in the plateau length, as also shown in Figure~\ref{Fig5}, can most likely be attributed to limitations in the writing accuracy using laser lithography or to inhomogeneities in the thin film.

\section{Conclusion} \label{Conclusion}
In this work, we investigated laser-lithographically written SNSPDs with widths between 400\,nm and 1.4\,$\upmu$m and lengths between 200\,$\upmu$m and 6\,mm. We demonstrated saturated detection efficiency at 775\,nm wavelength and 1550\,nm wavelength. The realization of laser-lithographically written SNSPDs enables easy and fast prototyping as well as structuring of large-scale SNSPD arrays, which we want to extend in future. In addition, micron-wide detectors can be used to drastically enhance the efficiency of integrated SNSPDs on low-index-contrast waveguides as UV-written waveguides in silica or titanium in-diffused wafeguides in lithium niobate.


\section*{Acknowledgements}
The study was partially funded by Bundesministerium f{\"u}r Bildung und Forschung (13N14911).

\section*{References}
\bibliographystyle{unsrt}
\bibliography{bibo}
\end{document}